\documentclass[journal]{IEEEtran}

\pdfoutput=1

\usepackage[noadjust]{cite}

\ifCLASSINFOpdf
   \usepackage[pdftex]{graphicx}
   \graphicspath{{./}}
   \DeclareGraphicsExtensions{.pdf,.jpeg,.png}
\else

\fi

\usepackage{amsmath,amsfonts,amssymb}
\usepackage{xcolor}

\newcommand{\x}{\mathbf{x}}
\newcommand{\m}{\mathbf{m}}

\newcommand{\F}{\mathbf{F}}
\newcommand{\dT}{\delta \mathbf{T}}
\newcommand{\dc}{\delta \mathbf{c}}
\newcommand{\G}{\mathbf{\Gamma}}
\newcommand{\Gn}{\G_{\text{noise}}}
\newcommand{\Gpr}{\G_{\text{prior}}}
\newcommand{\Gpo}{\G_{\text{post}}}
\newcommand{\R}{\mathbf{R}}
\newcommand{\V}{\mathbf{V}}
\newcommand{\D}{\mathbf{D}}
\newcommand{\Hess}{\mathbf{H}}

\newcommand{\norm}[1]{\left\lVert#1\right\rVert}

\begin{document}

\title{3D Wave-Equation-Based Finite-Frequency Tomography for
Ultrasound Computed Tomography}

\author{
\IEEEauthorblockN{Naiara~Korta~Martiartu\IEEEauthorrefmark{1},
        Christian~Boehm\IEEEauthorrefmark{1},
        and~Andreas~Fichtner\IEEEauthorrefmark{1}} \\
        
        \IEEEauthorblockA{\IEEEauthorrefmark{1}Institute of Geophysics,
        ETH Zurich, Zurich CH-8092, Switzerland}
        
\thanks{Correspondence to N. Korta Martiartu: naiara.korta@erdw.ethz.ch  }}%

\maketitle

\begin{abstract}

Ultrasound Computed Tomography (USCT) has great potential for 3D quantitative
imaging of acoustic breast tissue properties.
Typical devices include high-frequency transducers, which makes tomography
techniques based on numerical wave propagation simulations computationally
challenging, especially in 3D. Therefore, despite the finite-frequency nature of
ultrasonic waves, ray-theoretical approaches to transmission tomography are
still widely used.

This work introduces finite-frequency traveltime tomography to medical
ultrasound. In addition to being computationally tractable for 3D imaging at
high frequencies, the method has two main advantages: (1) It correctly accounts
for the frequency dependence and volumetric sensitivity of traveltime
measurements, which are related to off-ray-path scattering and diffraction. (2)
It naturally enables out-of-plane imaging and the construction of 3D images from
2D slice-by-slice acquisition systems.

Our method rests on the availability of calibration data in water, used to
linearize the forward problem and to provide analytical expressions of
cross-correlation traveltime sensitivity. As a consequence of the finite
frequency content, sensitivity is distributed in multiple Fresnel volumes,
thereby providing out-of-plane sensitivity. To improve computational efficiency,
we develop a memory-efficient implementation by encoding the Jacobian operator
with a 1D parameterization, which allows us to extend the method to large-scale
domains. We validate our tomographic approach using lab measurements collected
with a 2D setup of transducers and using a cylindrically symmetric phantom. We
then demonstrate its applicability for 3D reconstructions by simulating a
slice-by-slice acquisition systems using the same dataset.

\end{abstract}

\begin{IEEEkeywords}
Ultrasound computed tomography (USCT), finite-frequency tomography, Born
approximation, adjoint technique, breast imaging, resolution analysis,
point-spread function, traveltime
\end{IEEEkeywords}

\IEEEpeerreviewmaketitle

\section{Introduction}

Ultrasound computed tomography (USCT) is an emerging technique for diagnostic
imaging of breast tissue. In contrast to the commonly used sonography, USCT
uses both transmission and reflection data to provide quantitative images of
acoustic tissue properties. Empirical studies have demonstrated the diagnostic
value of this information, enabling non-invasive tissue characterization and
improving the specificity of standard imaging
modalities~\cite{Stavros1995,Greenleaf1997}.

Speed of sound, or velocity, is the most studied transmission property due to
its strong correlation with tissue density~\cite{Mast2000,SAK2017}, a risk
factor for breast cancer~\cite{Harvey2004}. Typical USCT systems use ray-based
tomographic algorithms to reconstruct this information from the first-arrival
times~\cite{Li2010,Maylin2015,GEMMEKE2017}.
These methods are robust and computationally efficient, making them very
attractive for clinical practice.
However, being based on the infinite-frequency assumptions, they neglect two
essential aspects of the traveltime measurements:
(1) In practice, traveltime differences are estimated by cross-correlating observed
signals and calibration data within a limited frequency band. Therefore, the measurements are inherently frequency-dependent \cite{Tong_1998}, meaning that results depend on the frequency content of the cross-correlated pulses.
(2) Due to scattering and diffraction, the traveltimes of the
finite-frequency waves are sensitive to tissue structure off the ray path. 
Consequently, the approximations of ray theory are only
valid for media with heterogeneities larger than the wavelength and the first
Fresnel volume~\cite{Williamson91}.

To improve the spatial resolution of the reconstructions, significant efforts
have been made to introduce imaging techniques that rely on more accurate
descriptions of wave propagation~\cite{Lavarello2013,Andre2013}.
This has led researchers to recognize similarities with seismology, and as a
result, seismic waveform tomography has been transferred to
USCT~\cite{Pratt07,Maylin2017,Boehm2018,Agudo2018}. Waveform tomography accounts for all
wave phenomena by numerically solving the wave equation. With sufficient data
coverage and quality, it can provide images with subwavelength spatial
resolution, however, at increased computational cost.
Applications to \textit{in vivo} data using slice-by-slice scanning systems have
shown very promising results~\cite{Pratt07,Sandhu2015}.
These studies use 2D approximations of the wave equation to provide
computationally tractable solutions. Therefore, artifacts may appear in the
images due to out-of-plane scattering and diffraction~\cite{Goncharsky16,Wiskin13,Sandhu2017}.
The practical adaptation of 3D waveform tomography in medical ultrasound
imaging still remains a challenge, mostly due to the high-frequency transducers
used in the acquisition systems. The number of wavelengths propagated through
the tissue exceeds 100~\cite{Pratt2018}, which makes waveform tomography both
computationally very expensive and highly sensitive to cycle-skipping and local
minima~\cite{Gauthier86,Bunks95}.

In this work, we present a novel approach to USCT that is computationally
tractable for 3D imaging at high frequencies and still capable of accounting for
first-order scattering and diffraction effects. This naturally enables
out-of-plane imaging, which makes it well suited, for instance, for the
construction of 3D images based on 2D slice-by-slice acquisition systems.
Originally proposed in
seismology~\cite{Luo91,Yomogida_1992,Maquering99,Dahlen2000}, numerous studies
have shown successful applications of finite-frequency tomography at
regional~\cite{Friederich_2003,Yoshizawa_2004,Sigloch_2008} and global
scales~\cite{Montelli04,Zhou_2006}. The method is valid for velocity contrasts
of up to 10\% with respect to a background model~\cite{Mercerat2012}, a
condition that is guaranteed in breast tissue. Furthermore, finite-frequency
considerations allow us to image velocity anomalies significantly smaller than
the first Fresnel zone~\cite{Baig2003,Joecker2006}.
This is especially true in the presence of adequate data coverage, which can
potentially be satisfied in USCT with rotations and translations of the
acquisition system. Our work therefore brings the method to the context of USCT
and honors the recent efforts for interdisciplinary collaboration between
seismology and USCT.

The tomographic method we present is based on the Born approximation of the wave
equation, and it uses adjoint
techniques~\cite{Tarantola_1988,Tromp2005,Fichtner_2006a,Plessix_2006} to
compute the Jacobian operator.
The latter relates velocity perturbations to traveltime observations that are
measured by the cross-correlation of ultrasonic signals with calibration data in
water. This results in a large-scale linear inverse problem. We propose a
cost-efficient implementation involving emitter-receiver geometry independent
one-dimensional parameterizations of the sensitivity kernels, which allows
massively parallelized operations on GPU architectures.

In what follows, we first introduce the theoretical aspects of the
finite-frequency tomography method, and we propose practical implementations in
the context of USCT. We then demonstrate its practical applicability using lab
data for both 2D and 3D reconstructions. Finally, we discuss the resolution
limits of the method, by relating it to the design of the acquisition systems.


%
%
%
%
%
%

%
%
%
%
%
%

%
%

\section{Finite-frequency traveltime tomography}

In USCT, we estimate the acoustic properties of breast tissue $\m$ from the
observations of the space- and time-dependent pressure wavefield $p(\x,t)$ at
transducer locations surrounding the breast. Observables and unknown parameters are related through
the acoustic wave equation, which is formulated, in the time domain and for
loss-less media, as
\begin{IEEEeqnarray}{c}
 \frac{1}{c^2(\x)} \partial^2_t 
 p(\x,t) - \rho(\x) \nabla \cdot \left(\frac{1}{\rho(\x)}
 \nabla p(\x,t)\right) = f(\x,t).
 \IEEEeqnarraynumspace
 \label{eq:Wave_equation}
\end{IEEEeqnarray}
Here, $f(\x,t)$ is the source term, and
the properties of the tissue are parameterized in terms of the velocity
$c(\x)$ and density $\rho(\x)$, being $\m = [c(\x); \rho(\x)]$. 
We denote by $p(\x_r,t;\x_s)$ the pressure field generated by an emitter at
$\x_s$ and recorded at receivers $\x_r$, and in general, we take specific
parts from this time series to define the observables.

Typically, density is assumed to be constant, and the velocity information is
retrieved from the first-arrival traveltimes. This can be formulated as a
least-squares problem, minimizing the sum of the squared residuals $\Delta
T(\x_r,\x_s;\m) = T_{\text{obs}}(\x_r,\x_s) - T(\x_r,\x_s;\m)$ between the
observed and predicted traveltimes for each emitter-receiver
combination, respectively. In practice, first-arrival
times are measured by the cross-correlation of the observed signals and calibration data. An
internally consistent way of formulating the optimization problem is therefore
by defining $\Delta T(\x_r,\x_s;\m)$ as the time shift $\tau$ where the cross-correlation
\begin{IEEEeqnarray}{c}
C(\tau) = \int_{t_0}^{t_1} p_{\text{obs}}(\x_r,t+\tau;\x_s)p(\x_r,t;\x_s,\m)dt
\label{cc_funct}
\end{IEEEeqnarray}
attains its maximum~\cite{Luo91}. Here, $[t_0,t_1]$ is the time
interval in which the first arrivals occur, and $p_{\text{obs}}$ and $p$ are the
observed and modelled ultrasonic signals, respectively.
If we drop the dependencies in $\x_r$ and $\x_s$ for clarity, $\Delta T$ satisfies
\begin{IEEEeqnarray}{c}
\left. \frac{dC(\tau)}{d\tau} \right\rvert_{\tau = \Delta T} =
\int_{t_0}^{t_1} \partial_t p_{\text{obs}} (t+\Delta
T) p(t;\m)dt = 0.
\label{relation_CC}
\end{IEEEeqnarray}
This equation establishes the implicit relation between $\Delta
T$ and $p(t;\m)$. For simplicity, we denote $\left. \partial_t p_{\text{obs}}
(t+\tau) \right\rvert_{\tau = \Delta T} = \partial_t p_{\text{obs}} (t+\Delta
T)$. In particular, for the velocity estimations, we
are interested in the relationship between the traveltime perturbations
$\delta T$ and velocity perturbations $\delta c$.
From \eqref{relation_CC}, by applying the chain rule for implicit
differentiation and integration by parts, we obtain
\begin{IEEEeqnarray}{c}
\delta T  = 
\frac{1}{N}\int_{t_0}^{t_1} \partial_t p_{\text{obs}}(t+\Delta T)\frac{\partial
p(t;\m)}{\partial c}\delta c \, dt,
\IEEEeqnarraynumspace
\label{forward1}
\end{IEEEeqnarray}
where
\begin{IEEEeqnarray}{c}
N = {\int_{t_0}^{t_1} \partial_t p_{\text{obs}}(t+\Delta
T)\partial_t p(t;\m)dt}.
\label{normalization}
\end{IEEEeqnarray}

The derivatives of the wavefield with respect to the model parameters that
appear in \eqref{forward1} can be computed from the Born approximation of
\eqref{eq:Wave_equation}. Assume we apply a perturbation in velocity
$c \rightarrow c + \delta c$ that causes a first-order perturbation in the
wavefield $p \rightarrow p + \delta p $. Then, the perturbed wavefield $\delta
p =  \frac{\partial p}{\partial c}\delta c$ is expressed, in terms of the
Green's function $G$, as (e.g.,~\cite{Tarantola84})
\begin{IEEEeqnarray}{c}
\delta p(\x,t) = \int_V
\int_{t_0}^{t_1} \frac{2 \delta c}{c^3(\x')} \partial^2_t
p(\x',t')G(\x,t;\x',t')dt' \, d\x',
\IEEEeqnarraynumspace
\label{wavefield_pert}
\end{IEEEeqnarray}
where $V$ is the volume of our ROI. By replacing
\eqref{wavefield_pert} in \eqref{forward1}, using the
reciprocity of the Green's function 
\begin{IEEEeqnarray}{c}
G(\x,t;\x',t') = G(\x',-t';\x,-t),
\end{IEEEeqnarray}
and expressing the wavefield at $\x_r$ as
\begin{IEEEeqnarray}{c}
p(\x_r,t) = \int_{V}
p(\x,t)\delta(\x-\x_r)d\x, 
\end{IEEEeqnarray}
we obtain
\begin{IEEEeqnarray}{c}
\delta T(\x_r,\x_s) = \int_{V} K(\x;{\x_r},{\x_s}) \delta
c(\x)\,  d\x.
\label{eq:forward_ffttt}
\end{IEEEeqnarray}
Here, $K(\x)= K(\x;{\x_r},{\x_s})$ is the cross-correlation traveltime
misfit sensitivity kernel defined as
\begin{IEEEeqnarray}{c}
K(\x) = \frac{-2}{
c^3(\x)} \int_{t_0}^{t_1} \partial_t p(\x,t;{\x_s})
\partial_t {p}^\dag(\x,t_1-t;{\x_r})dt.
\IEEEeqnarraynumspace
\label{eq:kernel_ffttt}
\end{IEEEeqnarray}
This is produced by the interaction between two
wavefields: the wavefield $p$ propagating forward from the emitters $\x_s$, and
the adjoint wavefield
\begin{IEEEeqnarray}{c}
p^{\dag}(\x,t) = \frac{1}{N} \int_{t_0}^{t_1}
\partial_t p_{\text{obs}}(\x_r,t'+\Delta T) G(\x,t';{\x_r},t)dt'
\IEEEeqnarraynumspace
\label{eq:adjoint_wavefield}
\end{IEEEeqnarray}
propagating backward in time from the receivers $\x_r$. 

Equation \eqref{eq:forward_ffttt} establishes the relationship between
cross-correlation traveltime shifts and velocity
perturbations in terms of the finite-frequency sensitivity
kernels $K(\x)$~\cite{Luo91,Maquering99,Dahlen2000}.
In principle, the sensitivity kernel in \eqref{eq:kernel_ffttt} depends on both
the observables and model parameters, and it is therefore a non-linear
relationship.
However, linearized approaches can be justified for velocity contrasts up to
10\% with respect to the background model~\cite{Mercerat2012}.

\subsection{Linearization} 

Assume that we represent the velocity distribution of the breast
tissue as $c(\x) = c_0 + \delta c(\x)$, where $c_0$ refers to a homogeneous
background model (water). If the differences are small
such that $|\delta c(\x)| \ll c_0$, the observed wavefield $p_{\text{obs}} =
p_0\, + \, \delta p$ will be a time shifted version of the background wavefield
$p_0$,
\begin{IEEEeqnarray}{c}
p_{\text{obs}}(\x_r,t+\Delta T)\approx p_0(\x_r,t).
\label{eq:Born_app} 
\end{IEEEeqnarray}

For a homogeneous unbounded medium, the 3D Green's function is given by
\begin{IEEEeqnarray}{c}
	G_0(\x_r,t;\x_s,t') = \frac{1}{4\pi R_{sr}}\delta \left(t
	-t'-\frac{R_{sr}}{c_0}\right),
\label{eq:Gree_homo}
\end{IEEEeqnarray}
where $R_{sr} = \norm{\x_s - \x_r}$, and assuming point sources, $f(\x,t) =
f(t)\delta(\x-\x_s)$, it follows that
\begin{IEEEeqnarray}{c}
	p_0(\x_r,t;\x_s) = \frac{1}{4\pi R_{sr} }f\left(t-\frac{R_{sr}}{c_0}\right).    
\label{eq:Wavefield_homo}
\end{IEEEeqnarray}

Upon inserting \eqref{eq:Born_app} - \eqref{eq:Wavefield_homo} into
\eqref{eq:kernel_ffttt}, we obtain the explicit expression of the linearized
traveltime sensitivity kernel $K_0$,
\begin{IEEEeqnarray}{rCl}
\IEEEyesnumber\label{eq:Linearized_sensitivity}
\IEEEyessubnumber*
	K_0 (\x) & = &
	\frac{A(\x)}{N}\int_{t_0}^{t_1} g_1(\x,t) g_2(\x,t)dt,
	\label{eq:Linearized_sensitivity_1}\\
	g_1(\x,t) & = & \partial_t
	f\left(t-\frac{R_{xs}}{c_0}\right), \\
	g_2(\x,t) & = & \partial^2_t f\left(t_1 - t -
	\frac{R_{xr}+R_{sr}}{c_0}\right), \\
	A(\x) & = & \frac{-R_{sr}}{2\pi {c_0}^3
	R_{xs} R_{xr}},\label{eq:Linearized_sensitivity_A}\\
	N & = & \int_{t_0}^{t_1}
	\partial_t f^2\left(t-\frac{R_{sr}}{c_0}\right) dt
	\label{eq:Linearized_sensitivity_end}, 
\end{IEEEeqnarray}
where $R_{xs} =\norm{\x-\x_s}$ and $R_{xr} =\norm{\x-\x_r}$.

\begin{figure*}[!t]
\centering
\includegraphics[width=6in]{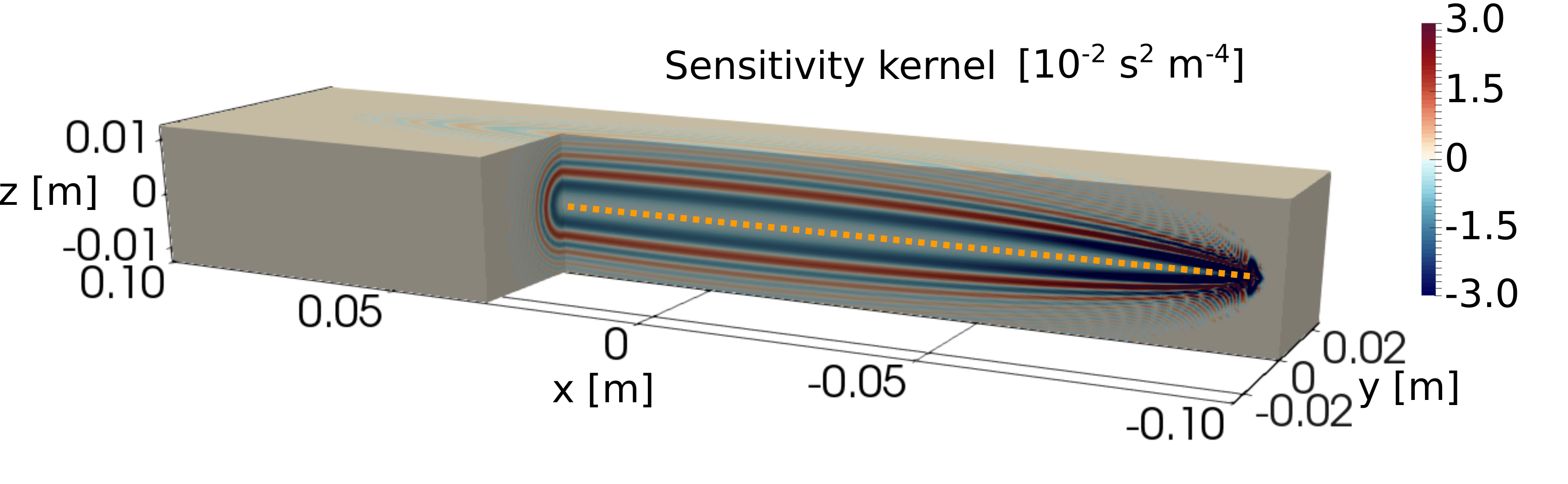}   
\caption{\label{fig:Kernel} Sensitivity kernel for a band-limited
signal with frequencies in the range of 1--3 MHz. The emitter and the receiver
are located at positions~(-0.095,0,0)~m and (0.095,0,0)~m, respectively. The
dashed line indicates the corresponding sensitivity predicted from ray theory.}
\end{figure*}

Equations \eqref{eq:Linearized_sensitivity_1} -
\eqref{eq:Linearized_sensitivity_end} define the key ingredient of the
forward operator of finite-frequency traveltime tomography, which does
neither depend on the unknown model parameters nor on the observed data.
Instead, it requires the source-time function $f(t)$, and this can be known, for
instance, from the calibration of the scanning device with measurements in
water.
An example of the finite-frequency sensitivity kernel for a band-limited pulse
with frequencies in the range of 1--3 MHz is shown in Fig.~\ref{fig:Kernel}.
Here, for comparison, we also illustrate the equivalent sensitivity predicted
from ray theory (dashed line). Unlike the ray-theoretical sensitivities, which
are confined to infinitesimally narrow paths, finite-frequency
sensitivities extend to finite volumes away from the geometrical ray. They
have an ellipsoidal shape defined by the Fresnel zones, with the strongest
contribution coming from the first Fresnel zone. Here, the negative sign
indicates that a positive velocity perturbation will produce earlier
first arrivals. For higher-order Fresnel zones, the sensitivities are
oscillatory with alternating signs, and their magnitude decrease due to
destructive interferences between the contribution of individual frequencies.
Contrary to ray theory, the sensitivities along the geometrical ray path are
zero, which is an effect of the cross-correlation measurements. The entire
waveform of the first arrivals contributes to the measured traveltimes, and it
is therefore the results of the interference between direct and scattered waves
\cite{Maquering99}. Higher frequencies produce narrower
sensitivity kernels, being consistent with what ray theory predicts. We refer
the reader to \cite{nolet_2008} for a more detailed discussion about the
sensitivity kernels.

\subsection{2D approximation} 

When the emitters and receivers are located in the same plane, 2D approximations
are often applied to accelerate the reconstructions \cite{Li2014}.
Assume we discretize the breast tissue using the Cartesian coordinates $\x =
[x;y;z]$ with the z-direction orthogonal to the acquisition plane. If the breast
tissue shows smooth variations of the velocity in z-direction, at least in the volumes defined by the sensitivity kernels, then $\delta c(\x) \approx
\delta c(x,y)$, and the forward problem can be reduced to 
\begin{IEEEeqnarray}{c}
\delta T(\x_r,\x_s) \approx \iint_{S} \left( \int_{z_0}^{z_1}
K_0(\x;{\x_r},{\x_s}) dz \right) \delta c(x,y) \, dx \, dy,
\IEEEeqnarraynumspace
\label{eq:forward_2D}
\end{IEEEeqnarray}
where $S$ is the ROI located at the acquisition plane, and $[z_0,z_1]$ is the
interval in which the main contribution of $K_0$ occurs. The accuracy of this
approximation will depend on both the structure of the breast tissue and the
frequencies used. At high frequencies, the sensitivity kernels extend to
narrower volumes, and the assumption of smooth variations can be better
justified.

\subsection{New parameterization of the forward operator}

The combination of \eqref{eq:forward_ffttt} and
\eqref{eq:Linearized_sensitivity}, or \eqref{eq:forward_2D} for 2D
approximations, describes the tomographic method presented in this study.
Using a compact notation, the linearized forward problem is written as
\begin{IEEEeqnarray}{c}
\dT = \F \dc,  
\label{eq:forward_compact}
\end{IEEEeqnarray}
where $\F \in \mathbb{R}^{M \times N}$ is the forward matrix, $\dT = [\delta
T_1;\ldots;\delta T_M]$, and $\dc = [\delta c_1;\ldots;\delta c_N]$. Here, $M$
and $N$ indicate the number of measurements and model parameters, respectively. 
Each row in $\F$ corresponds to 
a sensitivity kernel for an emitter-receiver pair. As Fig.~\ref{fig:Kernel} suggests, this
forward matrix is denser than its equivalent in ray theory. For large-scale
problems, the explicit computation of $\F$ may even become prohibitive due to
large memory requirements. We circumvent this by encoding the information
contained in the sensitivity kernels using a new parameterization. This allows
us to solve the inverse problem related
to~\eqref{eq:forward_compact} using iterative solvers, in which the matrix $\F$
is implicitly given through matrix-vector products.

The pattern shared by the sensitivity kernels can easily be observed in
the temporal Fourier domain. We first convert \eqref{forward1}
and \eqref{normalization} with the
approximation made in \eqref{eq:Born_app} \cite{Tong98,nolet_2008}:
\begin{IEEEeqnarray}{c}
\delta T = - \frac{\operatorname{Re} \int_{0}^{\infty} i \omega
p(\omega)^*\delta p(\omega) d\omega}{\int_{0}^{\infty} \omega^2
p(\omega)^* p(\omega)d\omega}.
\label{eq:forward1_freqdomain}
\end{IEEEeqnarray}
Here, $p(\omega)$ denotes the Fourier-transformed pressure field, with
angular frequency $\omega$, and we omitted the spatial dependency for
clarity. The superscript $*$ and $\operatorname{Re}$ denote the complex
conjugate and the real part of the complex number, respectively.

For a point source $f(\x,\omega) = f(\omega)\delta(\x-\x_s)$, the pressure
fields at receiver locations $\x_r$ can be expressed in terms of Green's
functions as
\begin{IEEEeqnarray}{c}
p(\omega)= f(\omega)G_0(\x_r,\omega;\x_s),
\label{eq:pressure_field1}
\end{IEEEeqnarray}
\begin{IEEEeqnarray}{c}
\delta p(\omega)= - \int_{V} \frac{2 \omega^2 \delta c(\x')}{c_0^3}
p(\x',\omega;\x_s)G_0(\x_r,\omega;\x') d\x',
\IEEEeqnarraynumspace
\label{eq:pressure_field2}
\end{IEEEeqnarray}
where the frequency-domain Green's function is
\begin{IEEEeqnarray}{c}
G_0(\x_r,\omega;\x_s)= \frac{1}{4\pi R_{sr}}
\exp\left(-i\frac{\omega}{c_0}R_{sr}\right).
\IEEEeqnarraynumspace
\label{eq:Green_freq}
\end{IEEEeqnarray}
Upon inserting \eqref{eq:pressure_field1} - \eqref{eq:Green_freq} into
\eqref{eq:forward1_freqdomain}, the
sensitivity kernel $K_0 = K_0(\x)$ takes the form
\begin{IEEEeqnarray}{c}
K_0 = A \frac{\int_{0}^{\omega} \omega^3 |f(\omega)|^2
\sin\left(\frac{\omega}{c_0}(R_{xs}+R_{xr}-R_{sr})\right) d\omega
}{\int_{0}^{\omega} \omega^2 |f(\omega)|^2 d\omega}.
\IEEEeqnarraynumspace
\label{eq:Kernel_freq}
\end{IEEEeqnarray}
The term $A = A(\x)$ is the same as in \eqref{eq:Linearized_sensitivity_A}, and
it mostly accounts for geometrical spreading. Because it does not depend on $\omega$,
it can be included in the sensitivity kernels at a later stage. We
therefore do not consider it for the new parameterization of the forward
operator.

\begin{figure}[!t]
\centering
\includegraphics[width=3.45in]{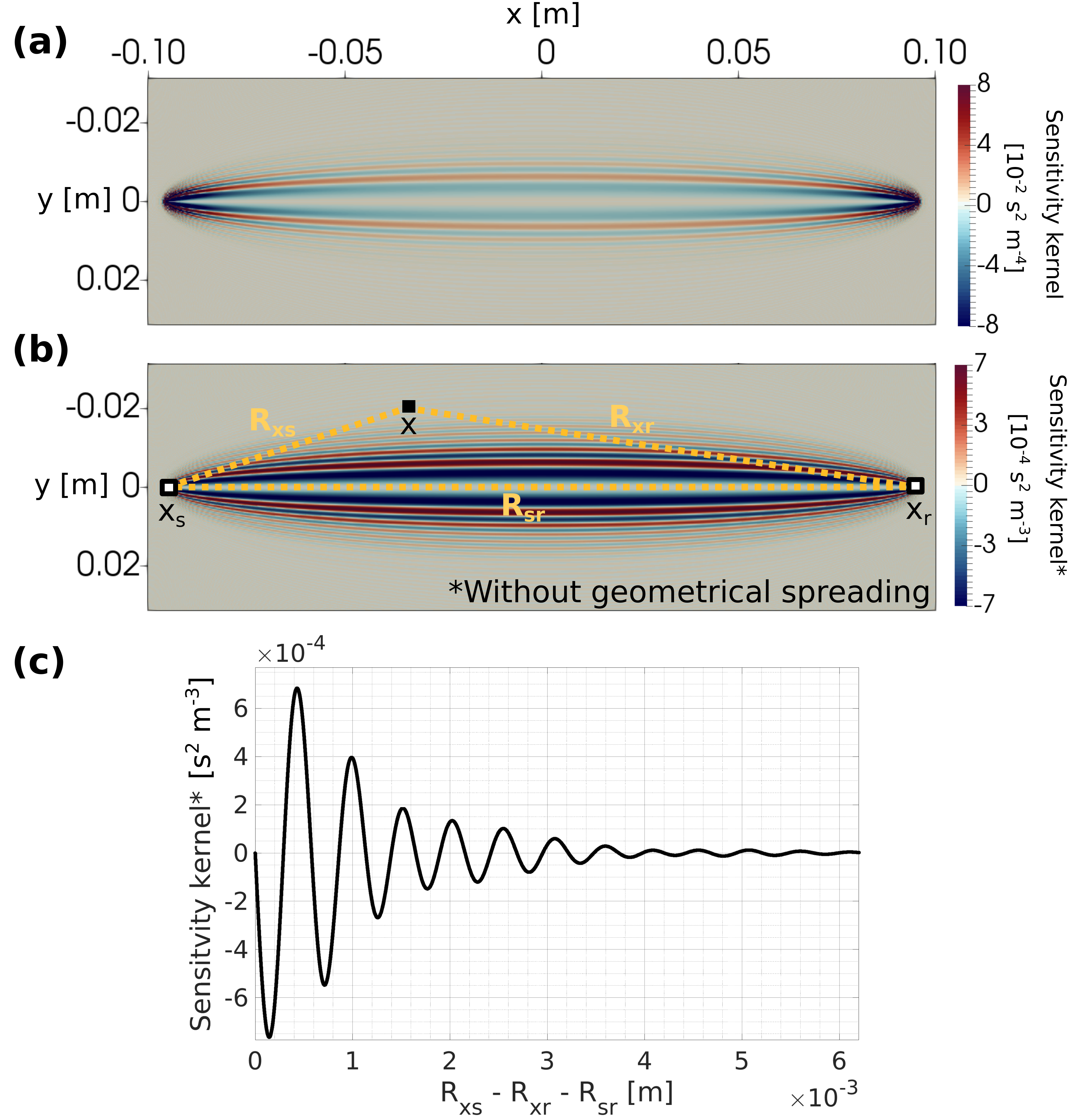}   
\caption{\label{fig:Pattern} Cross-section in xy-plane of the sensitivity
kernel in Fig.~\ref{fig:Kernel} (a) with and (b) without the geometrical
spreading term $A(\x)$. (c) Sensitivity kernel in (b) represented as a
function of $R = R_{xs}+R_{xr}-R_{sr}$. These distances are indicated in (b) with dashed
lines, together with the positions of the emitter $\x_s$, receiver $\x_r$ and
an arbitrary spatial location $\x$.}
\end{figure}

From \eqref{eq:Kernel_freq}, we see that by defining $K_0$ in terms of
${R=R_{xs}+R_{xr}-R_{sr}}$, the sensitivities for any emitter-receiver
combination are represented by the same analytical function. We show an example of this in Fig.
\ref{fig:Pattern}. The function describes the diffraction pattern observed in
the sensitivity kernels, that is, the values and locations of the Fresnel zones.
Being independent of the emitter-receiver geometry, it essentially encodes the
complete forward operator $\F$, without the need of computing it explicitly.
Usually, we need very few terms to represent
accurately the function in \eqref{eq:Kernel_freq}, and the sparsity of the
forward operator can be controlled by truncating the maximum value of the parameter $R$. We therefore
store the coefficients and arguments of the involved trigonometric functions,
and we compute the elements of $\F$ on the fly through the matrix-vector
products required during the iterative linear inversion. For each matrix-vector operation, the actual values
of the sensitivities are computed by evaluating the analytical function in the
discretization grid, and multiplying them by the corresponding geometrical
spreading term $A(\x)$. These computations can be done very efficiently in GPU
architectures, for which trigonometric functions are optimized operations.

\subsection{Inversion and resolution analysis}

The inverse problem can be formulated as a least-squares minimization, where we
estimate the model parameters that minimize the objective functional
\begin{IEEEeqnarray}{c}
\chi(\dc) = \frac{1}{2}{\norm{\F\dc - \dT_{\text{obs}}}_{\Gn^{-1}}^2} + \alpha
J(\dc).
\label{eq:leastSquare_misfit1}
\end{IEEEeqnarray}
Here, the first term accounts for the discrepancies between the predicted and
observed traveltimes, in which we assume Gaussian noise with zero mean and
covariance matrix $\Gn$, and the weighted norms are defined as
$\norm{\x}^2_{\G^{-1}} = \x^{T}\G^{-1}\x$. $J(\dc)$ is the regularization term,
and $\alpha > 0$ is the regularization parameter that balances the contribution of both
terms. We solve the regularized least squares
problem~\eqref{eq:leastSquare_misfit1} iteratively and without
explicitly constructing the forward matrix $\F$.

The regularization term incorporates our prior information about the
model parameters to mitigate the ill-posed inverse problem, and to ensure
meaningful solutions. If the prior is Gaussian with covariance matrix $\Gpr$,
and ${J(\dc) = \frac{1}{2}\norm{\dc}_{\Gpr^{-1}}^2}$, the solution can be
analytically estimated as
\begin{IEEEeqnarray}{c}
\dc_\text{est} = (\F^{T}\Gn^{-1}\F + \alpha \Gpr^{-1})^{-1} \F^{T} \Gn^{-1} \dT.
\label{eq:leastSquare_misfit}
\end{IEEEeqnarray}
Uncertainties in the solution \eqref{eq:leastSquare_misfit} are described
by means of the posterior covariance 
\begin{IEEEeqnarray}{c}
\Gpo = (\F^{T}\Gn^{-1}\F + \alpha \Gpr^{-1})^{-1},
\label{eq:posterior}
\end{IEEEeqnarray}
and model resolution matrix
\begin{IEEEeqnarray}{c}
\R = (\F^{T}\Gn^{-1}\F + \alpha \Gpr^{-1})^{-1} \F^{T}\Gn^{-1}\F
\label{eq:resolution}
\end{IEEEeqnarray}
that satisfies $\dc_\text{est} = \R \dc_\text{true}$. 

For large-scale problems, even when $\F$ is explicitly available, computing
$\Gpo$ and $\R$ may be challenging. However, we can benefit from the rapidly
decaying eigenvalues of $\Gpr^{1/2}\F^{T}\Gn^{-1}\F\Gpr^{1/2}$ to compute their
low-rank approximations~\cite{Flath2011}. We do this by retaining the $k < N$
largest eigenvalues and the corresponding eigenvectors. That is, we approximate 
\begin{IEEEeqnarray}{c}
\Gpr^{1/2}\F^{T}\Gn^{-1}\F\Gpr^{1/2} = \V \mathbf{\Lambda} \V^T \approx \V_k
\mathbf{\Lambda}_k \V_k^T,
\label{eq:low_rank}
\end{IEEEeqnarray}
where $\mathbf{\Lambda} = \text{diag}(\lambda_i) \in \mathbb{R}^{N \times N}$
and $\V \in \mathbb{R}^{N \times N}$ are the eigenvalue and eigenvector matrix,
respectively.
Then, the low-rank approximations of $\Gpo$ and $\R$ are given by
\begin{IEEEeqnarray}{c}
\Gpo \approx  \alpha^{-1} \Gpr^{1/2} \left(\mathbf{I} -\V_k \D_k
\V_k^T\right)\Gpr^{1/2} ,
\label{eq:posterior_app}
\end{IEEEeqnarray}
\begin{IEEEeqnarray}{c}
\R \approx \Gpr^{1/2}\V_k \D_k \V^T_k  \Gpr^{-1/2}.
\label{eq:resolution_app}
\end{IEEEeqnarray}
with $\D_k = \text{diag}\left(\frac{\lambda_i}{\lambda_i + \alpha}\right) \in
\mathbb{R}^{k \times k}$. For a more detailed derivation of
\eqref{eq:posterior_app} and \eqref{eq:resolution_app}, the reader is referred
to~\cite{Flath2011}.

The posterior covariance $\Gpo$ is useful to interpret the reliability of our
velocity estimations. The diagonal entries indicate the variances of the
individual parameters, and the off-diagonal entries show the correlations
between the errors in different model parameters. As observed in
\eqref{eq:posterior_app}, $\Gpo$ is the result of extracting from our prior
uncertainties the information we gain from the data.

The resolution matrix $\R$ indicates how well the model parameters are resolved
in the inversion. It assumes that the observed data fully satisfy the
forward problem, and when the model parameters are perfectly resolved, $\R$
equals the identity matrix. In general, however, $\R \neq \mathbf{I}$, and
the estimated parameters are weighted averages of the true parameters. Each column
of $\R$ is interpreted as a point-spread function (PSF), which illustrates the
blurring of one parameter into others. PSFs are therefore useful to describe how
independently an individual parameter can be resolved by the data.
 
For high-dimensional problems, the low-rank approximation of $\R$ may still be
prohibitively expensive to compute. Alternatively, we can estimate PSFs from the
Hessian operator of the misfit term in \eqref{eq:leastSquare_misfit},
$\Hess_{\text{misfit}} = \F^{T}\Gn^{-1}\F$ \cite{Fichtner2015}. From the
definition of $\R$, we observe that
\begin{IEEEeqnarray}{c}
\dc_{\text{est}} = \R \dc_{\text{true}} = (\Hess_\text{misfit} + \alpha
\Gpr^{-1})^{-1}\Hess_{\text{misfit}}\dc_{\text{true}}.
\IEEEeqnarraynumspace
\label{eq:resolution_2}
\end{IEEEeqnarray}
Here, the term $\Hess_{\text{misfit}}\dc_{\text{true}}$ indicates the direction
of the model updates. When this is multiplied by $(\Hess_\text{misfit} +
\alpha\Gpr)^{-1}$, we converge to the solution in a single iteration. 
If we apply $\Hess_{\text{misfit}}$ to a point-localized model
perturbation $\dc$, it will therefore provide a conservative
estimation of the PSFs. In our study, we use this approach for the resolution
analysis of the finite-frequency traveltime tomography, and in particular, to
understand how the vertical resolution is related to the acquisition design.

\section{2D lab data application}

\begin{figure}[!t]
\centering
\includegraphics[width=3.45in]{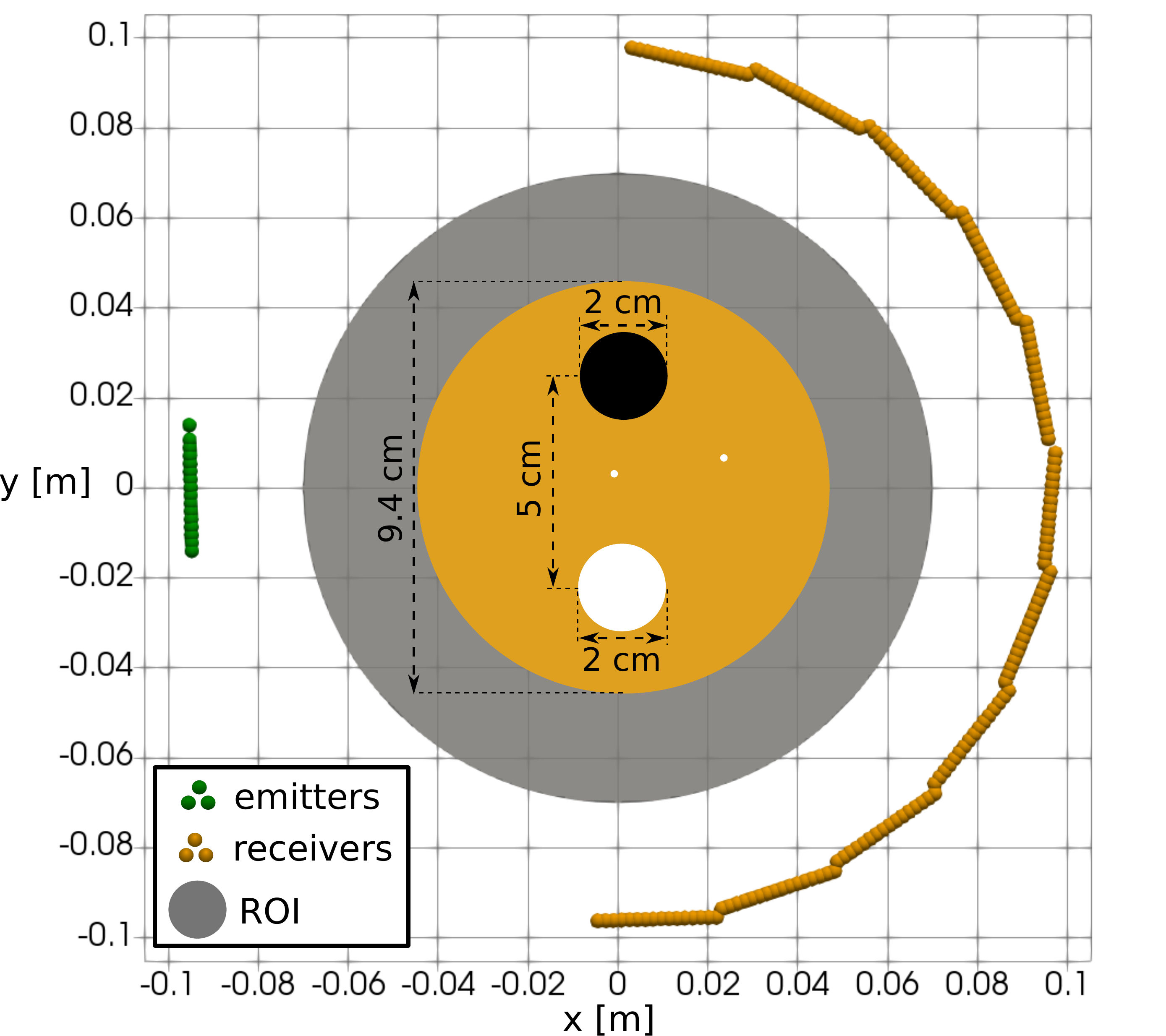}   
\caption{\label{fig:Setup} Acquisition system and an illustration of the tissue
mimicking phantom used for the lab measurements. Here the colors do not have a
quantitative meaning. }
\end{figure}

We consider the dataset provided by the Spanish National Research Council (CSIC)
and the Complutense University of Madrid (UCM) as part of the \textit{SPIE USCT
Data Challenge 2017}~\cite{Ruiter2018,Camacho2012}. The experimental setup used
for the measurements is shown in \figurename~\ref{fig:Setup}. The acquisition
system consists of two 16-element linear transducer arrays.
The elements in one array act as emitters with a central frequency of 3.2~MHz
and bandwidth of 50\%, while the others are receiving. To acquire transmission
data, the receiving array is placed in 11 different positions per position of
the emitting array. The whole system is rotated 23 times with respect to the
vertical axis describing a circle of 95~mm radius and providing a total of 64768
A-scans. The maximum data coverage is obtained in a circular ROI of 70~mm
radius, indicated in gray in \figurename~\ref{fig:Setup}. The same figure also
illustrates the cylindrical phantom used for the measurements. It is based on
water, gelatine, alcohol and graphite powder, and it includes an homogeneous
background of 94~mm diameter, two inclusions of 20~mm diameter, and two steel
needles of 0.25~mm diameter. The phantom is submerged in water with a calibrated
velocity of 1479.7~m/s.

\begin{figure}[!t]
\centering
\includegraphics[width=3.45in]{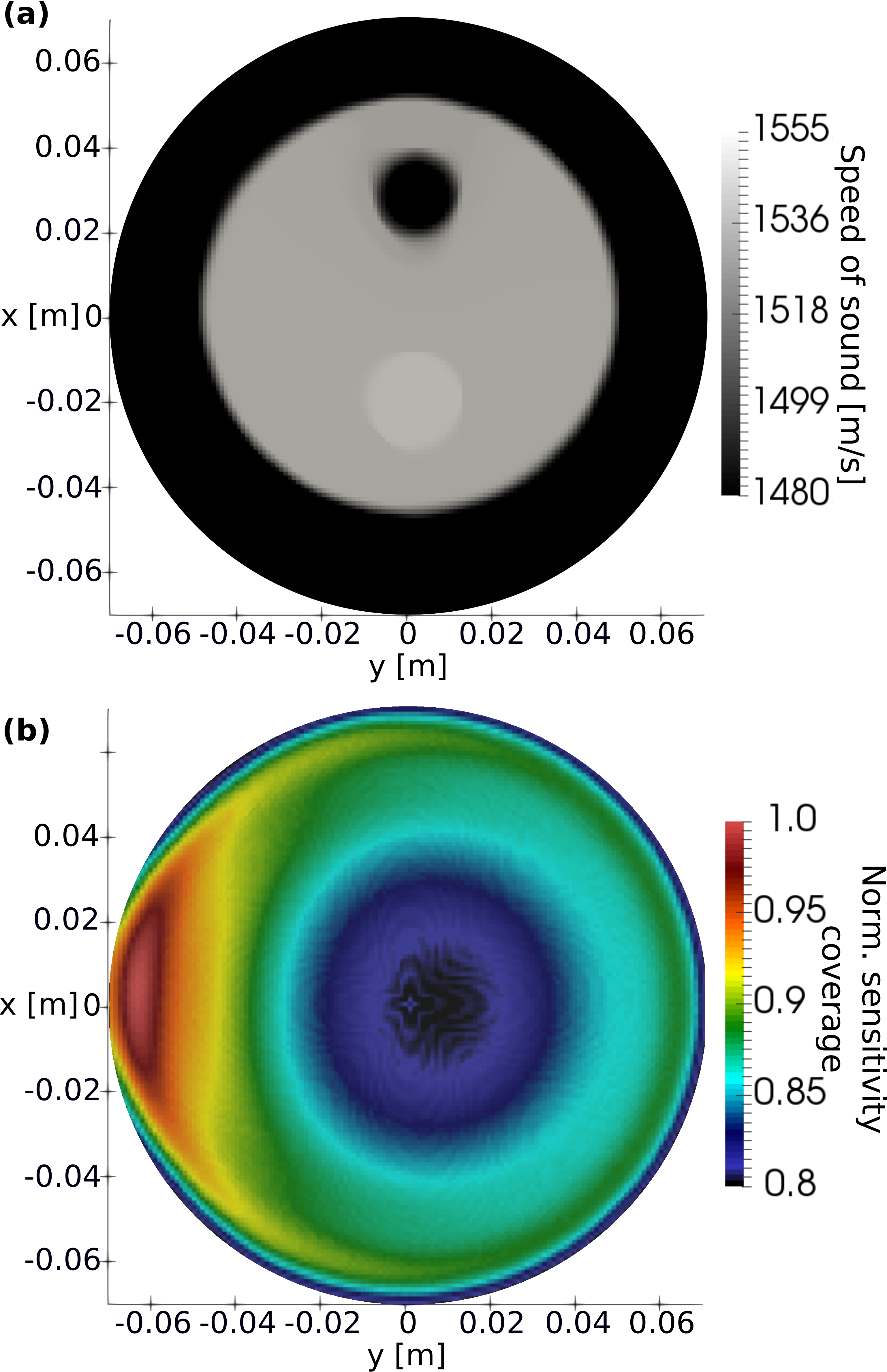}     
\caption{\label{fig:2Drec}(a) 2D velocity reconstruction and (b) normalized
sensitivity coverage. Both images only show the ROI indicated in gray in \figurename~\ref{fig:Setup}.}
\end{figure}

\begin{figure}[!t]
\centering
\includegraphics[width=3.45in]{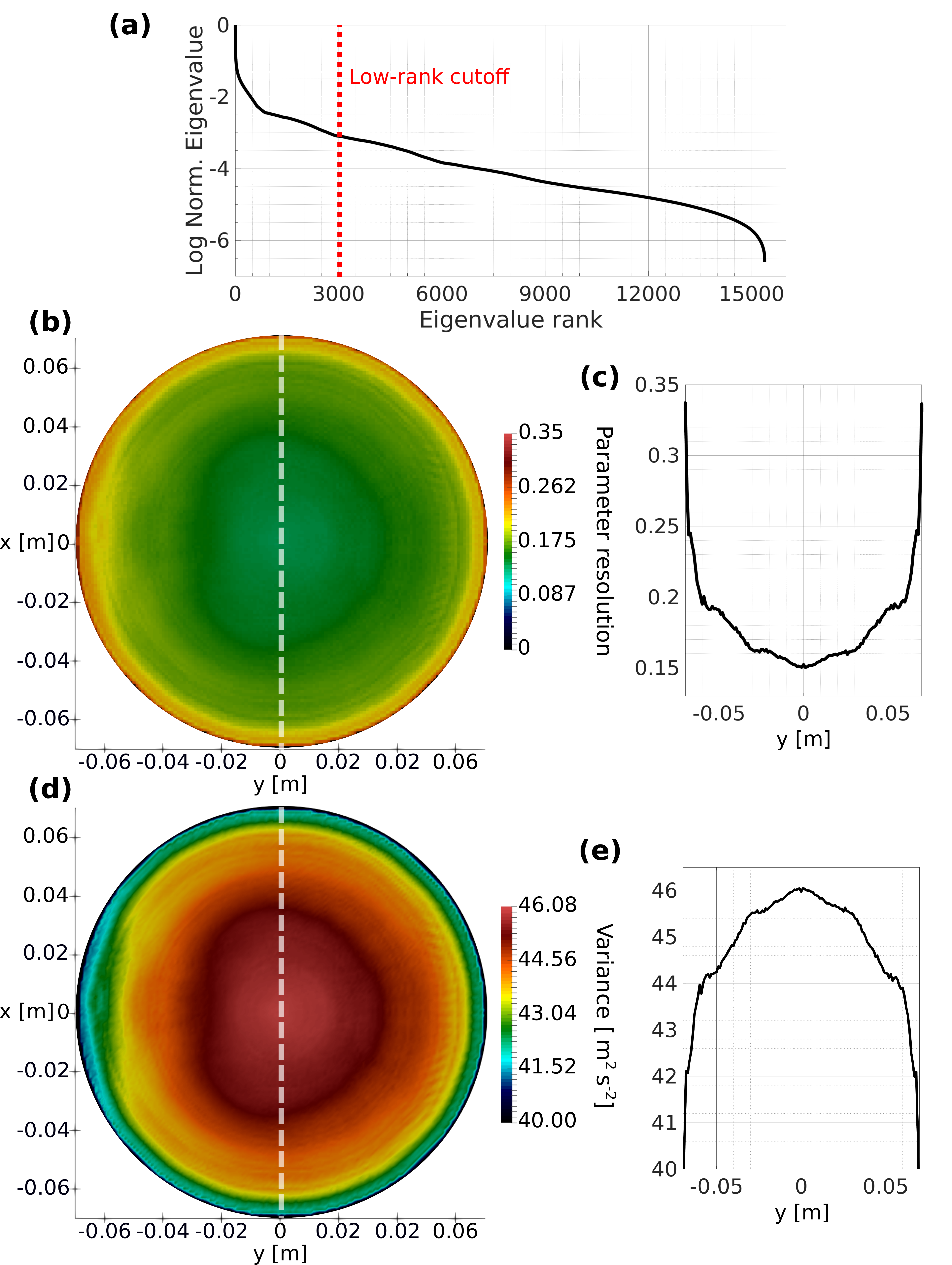}
\caption{\label{fig:2DResolution} (a) Singular value spectrum of the 2D forward
operator. The red line indicates the low-rank cutoff
used to approximate the resolution and posterior covariance
matrices. (b), (c) Diagonal elements of the resolution matrix and a cross
section at the position indicated by the dashed white line, respectively. (d),
(e) Diagonal elements of the posterior covariance matrix and a cross section at
$y=0\,\text{m}$, respectively.
}
\end{figure}

Although the experiment is inherently 3D, the lack of vertical variation in
acoustic properties of the phantom makes 2D approximations reasonable.
\figurename~\ref{fig:2Drec}(a) shows the velocity reconstruction obtained using
the 2D forward problem in~\eqref{eq:forward_2D}.
Here, and hereafter, we apply total variation regularization~\cite{Jensen2012},
and we discretize the model parameters using a rectilinear grid with 1~mm mesh
size. This gives a total of 15373 unknowns in the ROI, and thus, a forward
operator with dimensions $64768 \times 15373$. Despite the over-determined
nature of the problem, redundancies exist due to imperfect data coverage, and
the inverse problem is ill-posed (see \figurename~\ref{fig:2DResolution}(a)). By
summing the absolute values of the rows in the forward operator, we also compute
the sensitivity coverage of the experiment, shown in
\figurename~\ref{fig:2Drec}(b). We observe that the coverage decreases towards
the center of the ROI, which is a consequence of the approximately regular
distribution of the transducers~\cite{Gemmeke2014,Korta2017}. The region with
highest coverage in the left side is caused by the overlapped positions of
transducers in the last rotation of the acquisition system.

Our reconstruction recovers accurately both velocity heterogeneities and the
homogeneous background of the phantom. First arrival traveltimes do not contain
information about the needles, which act as scatterers, and they are therefore
invisible for transmission tomography.
Empirical velocity measurements of the true phantom are not available, and this
excludes a quantitative assessment of our velocity estimations. Yet, our
results are in agreement with the reconstructions obtained by other
groups~\cite{Ruiter2018}, which certainly shows the efficiency of our imaging
method. Whereas both inclusions have the same size, a closer look to our
reconstruction reveals that the high velocity inclusion is slightly bigger than
the low velocity one. This may be an effect of our linearization approximation, in
which the sensitivity kernels are computed in an homogeneous model, and
therefore, they neglect the bending that occurs in heterogeneous media.

To assess the quality of our solution, a comparison between the true
model and the solution might be insufficient. Although this suggests that the main
features of the true phantom are well resolved, it obscures the actual contribution of our prior
knowledge introduced by the choice of the regularization. For this example a
comprehensive analysis of the resolution is available. We compute the singular
value spectrum of the forward operator, shown in
\figurename~\ref{fig:2DResolution}(a). As we observe, the singular values decay
rapidly in magnitude, suggesting that the data only contain information about few effective model parameters. We truncate the singular value spectrum after a
decrease of three orders of magnitude and consider values below as the effective
nullspace.
We therefore retain 3000 singular values, and we compute low-rank
approximations of the posterior covariance and resolution matrices, see~\eqref{eq:posterior_app}
and~\eqref{eq:resolution_app}, respectively. Here, we assume equally
reliable observations, i.e., $\Gn = \sigma_p^{-2} \mathbf{I}$ with $\sigma_p
= 2.5 \cdot 10 ^{-8}~\text{s}$ being the minimum cross-correlation time
shift that can be measured.
The diagonal elements of these matrices are shown in Figs.
\ref{fig:2DResolution}(b) and \ref{fig:2DResolution}(d), respectively. To better
understand the images, we moreover show their cross sections at $y=0\,\text{m}$ in Figs.
\ref{fig:2DResolution}(c) and \ref{fig:2DResolution}(e).
For both quantities we observe similar features: the parameter resolution
decreases gradually, and the variances increase towards the center of the ROI.
This means that we are more uncertain about the parameters resolved at the
center of the ROI, which corresponds to the observations made in the
sensitivity coverage.

\begin{figure*}[!t]
\centering
\includegraphics[width=7in]{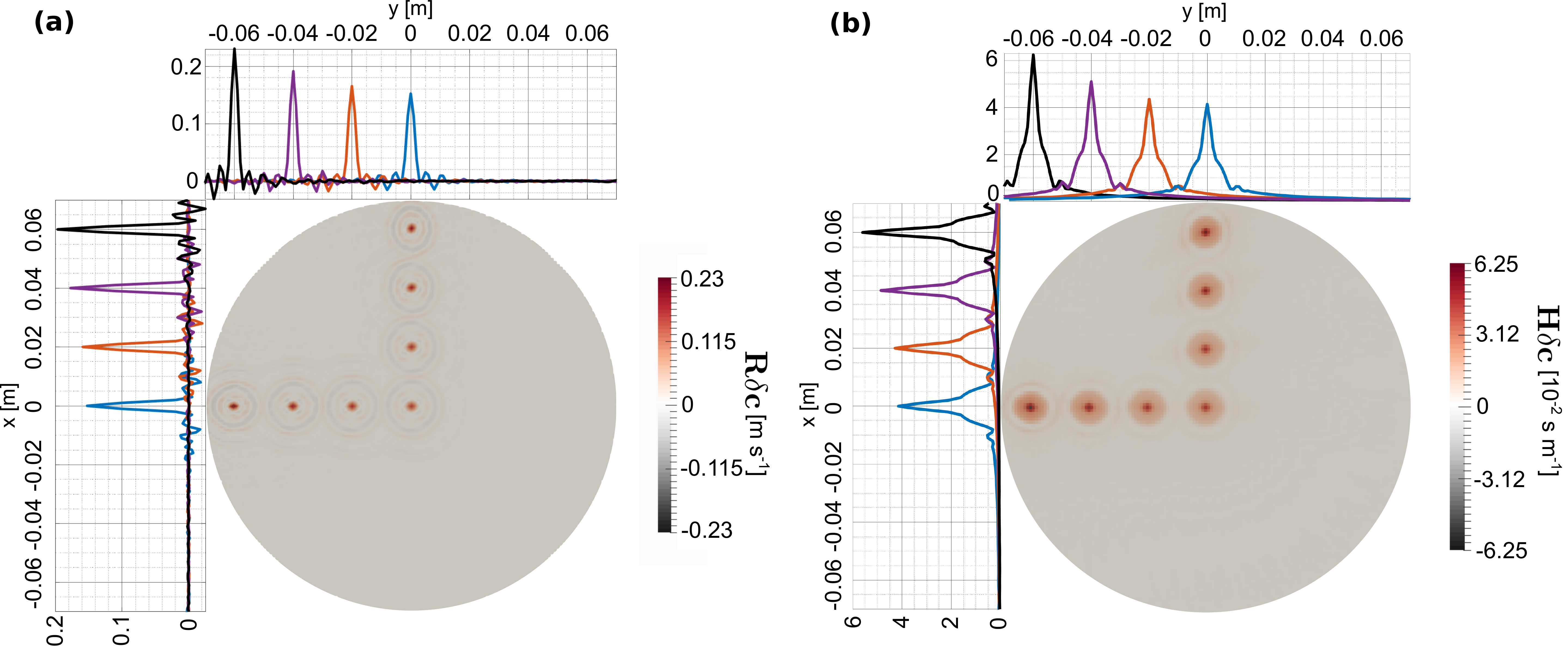}
\caption{\label{fig:PSF2d} PSFs corresponding to
point-localized unit perturbations at positions
$(x,y)=(0,0),\,(0,2),\,(0,4),\,(0,6),\,(-6,0),\,(-4,0),\,(-2,0)\,\text{cm}$.
These are estimated using (a) the resolution matrix, and (b) Hessian-vector products.
In both figures we show the horizontal and vertical cross sections
at $y=0\,\text{m}$ and $x=0\,\text{m}$, respectively.}
\end{figure*}

The diagonal elements of the resolution matrix are useful to understand where we
may expect better resolved parameters. However, it misses useful information
encoded in the off-diagonal elements about trade-offs and spatial
correlations between parameters that occur due to imperfect data coverage. In
\figurename~\ref{fig:PSF2d}(a), we represent few columns of the resolution
matrix, which describe the blurring effect introduced by the inverse operator. These are the PSFs
associated to point-localized unit perturbations at positions
$(x,y)=(0,0),\,(0,2),\,(0,4),\,(0,6),\,(-6,0),\,(-4,0),\,(-2,0)\,\text{cm}$.
Our resolution analysis reveals that the
smearing mainly extends to circular areas of 1 mm
radius. The worst resolution is obtained at the center of ROI, and
the trade-offs decrease in spatial extension towards its boundary. Similar
behaviour is observed in the amplitude of the PSFs. 

The information extracted from the resolution matrix corresponds to an idealized
situation, and it may lead us to rather optimistic conclusions. It assumes that
(1) the observed data satisfies exactly the forward problem
in~\eqref{eq:forward_compact}, and (2) our model estimation fully converges to
the solution.
For 3D problems the computation of the resolution matrix becomes prohibitive,
and therefore, PSFs can only be estimated through Hessian-vector products
(see~\eqref{eq:resolution_2}). These quantities, which indicate the direction of
the single-iteration model update, are considered as conservative estimations of
PSFs \cite{Fichtner2011,Fichtner2015}. Here we compare both results in order to
gain deeper understanding for the following sections.

The estimations of the PSFs using Hessian-vector products are shown in
\figurename~\ref{fig:PSF2d}(b) for the same locations as before. As expected,
the parameter trade-offs extend to wider areas than the ones estimated using the
resolution matrix. This may be a consequence of the projection of the 3D
sensitivities to the x-y plane, which is intrinsic to our definition of the 2D
forward problem (see~\eqref{eq:forward_2D}). Although the trade-offs are mainly
localized in a circle of 5 mm radius around the positions of the perturbations,
the locations of the most significant values agree with our observations in
\figurename~\ref{fig:PSF2d}(a). In fact, the principal difference between
Figs.~\ref{fig:PSF2d}(a) and \ref{fig:PSF2d}(b) is due to the normalization
factor in \eqref{eq:resolution_2}. PSFs estimated by Hessian-vector products are
dominated by smoother eigenvectors associated to largest eigenvalues.
Consequently, they obscure the small-scale features provided by other
eigenvectors that the inverse operator resolves.

\section{3D reconstruction with 2D lab dataset} 

Lab experiments are 3D in nature, and volumetric reconstructions may be
preferable, especially when the observations are sensitive to regions with
vertically varying structure. In this section, we use the same dataset as in the
example before to illustrate the potential of our method to image out of
plane.

\begin{figure*}[!t]
\centering
\includegraphics[width=7in]{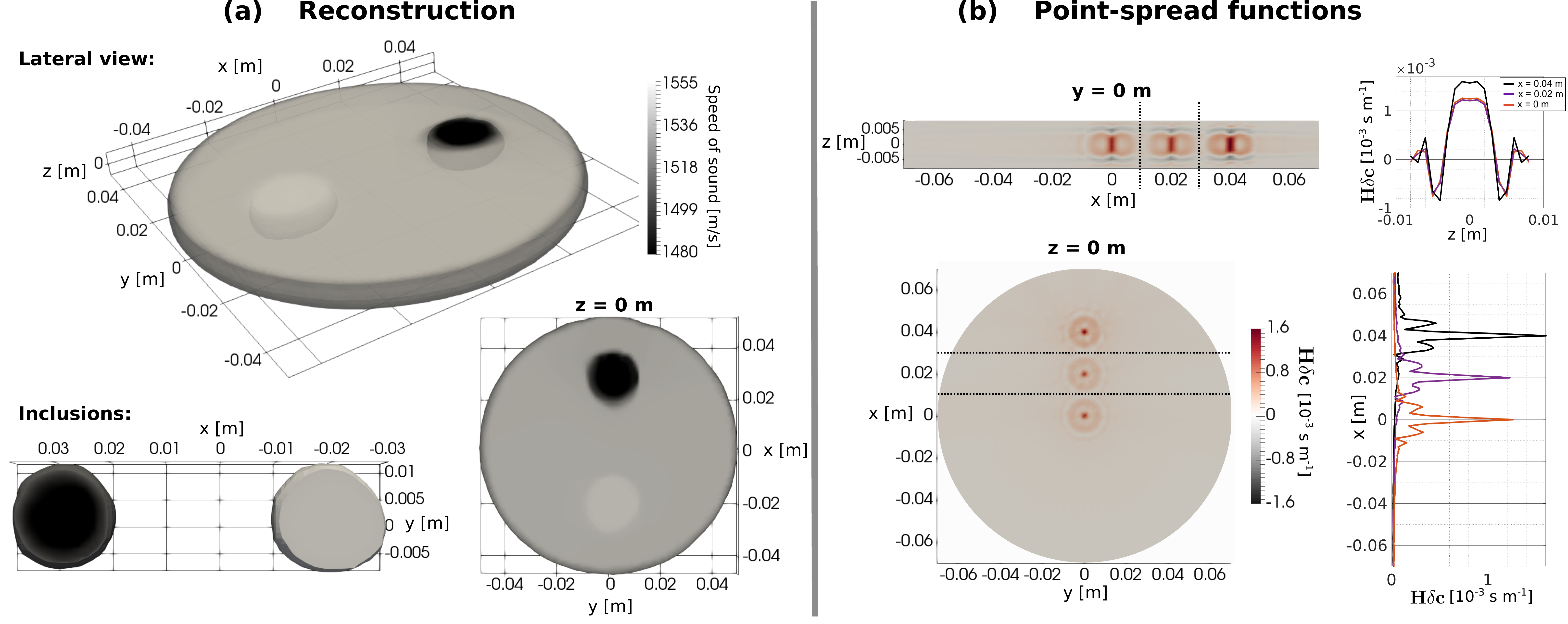}
\caption{\label{3Drec_1elev}(a) 3D velocity reconstruction using the same
dataset as in the 2D example. We show the lateral view of the volumetric
reconstruction, the top view of the isolated inclusions, and a cross section at
$z = 0\,\text{m}$. We reduce the opacity of the first image to visualize the
inclusions.
(b) Estimations of PSFs due to unit perturbations at locations
$(x,y,z)=(0,0,0),\,(2,0,0),\,(4,0,0)\,\text{cm}$. We show the 2D and 1D cross
sections in vertical and horizontal directions. Dashed lines indicate that PSFs are
computed individually, although we visualize all in the same image.}
\end{figure*}

We apply the forward problem in~\eqref{eq:forward_ffttt} to reconstruct the 3D
velocity distribution. The result is shown in \figurename~\ref{3Drec_1elev}(a),
and it proves that indeed finite-frequency tomography is capable of providing
volumetric images from 2D acquisition systems. In general, our estimated
velocity model recovers the main features of the true phantom, with similar
accuracy as in the 2D example.
The vertical width of the reconstruction corresponds to the region with the
highest sensitivity coverage.
This is controlled by the interaction of the first Fresnel zone of the
sensitivities corresponding to each emitter-receiver combination. In this
application, the maximum Fresnel width is $\sqrt{\frac{c_0 L_{\text{max}}}{f_c}}
= 9.4\,\text{mm}$, where $c_0 = 1479.7~\text{m/s}$ is the background water
velocity, $L_{\text{max}}=0.19~\text{m}$ is the maximum emitter-receiver
distance, and $f_c = 3.2~\text{MHz}$ is the dominant frequency of the emitting
signal.
The vertical thickness of our reconstruction will therefore be constrained by
this value.

To better understand the quality of our reconstruction, we compute a collection
of PSFs estimated by Hessian-vector products. These are
visualized in \figurename~\ref{3Drec_1elev}(b) as vertical and horizontal cross
sections.
The results reveal a similar horizontal resolution as in the 2D example, but a
poor vertical resolution that essentially makes velocity parameters
indistinguishable in z-direction. As mentioned before, its extent comprises the
most covered region of approximately 8~mm width, and it remains constant along
the horizontal direction. In this example, the lack of crossing sensitivity
kernels in z-direction decreases our ability to constrain independently the
parameters.

\section{Simulated full 3D experiment: slice-by-slice acquisition}

To improve the vertical resolution of an experiment, it is essential to provide
measurements with crossing or overlapping sensitivities. For the previous
experiment, this can be done, for instance, by collecting additional
measurements at different elevations. Because the phantom has cylindrical
symmetry, we simulate a slice-by-slice acquisition, and we assume that the same
data have been recorded at different elevations. The aim of this example is twofold:
on one hand, we want to show that finite-frequency tomography is a powerful tool
to provide consistent full 3D images from slice-by-slice acquisition systems;
and on the other hand, we want to investigate the conditions for a meaningful
vertical resolution. We relate the latter to the vertical spacing between
different elevations.

\begin{figure*}[!t]
\centering
\includegraphics[width=7in]{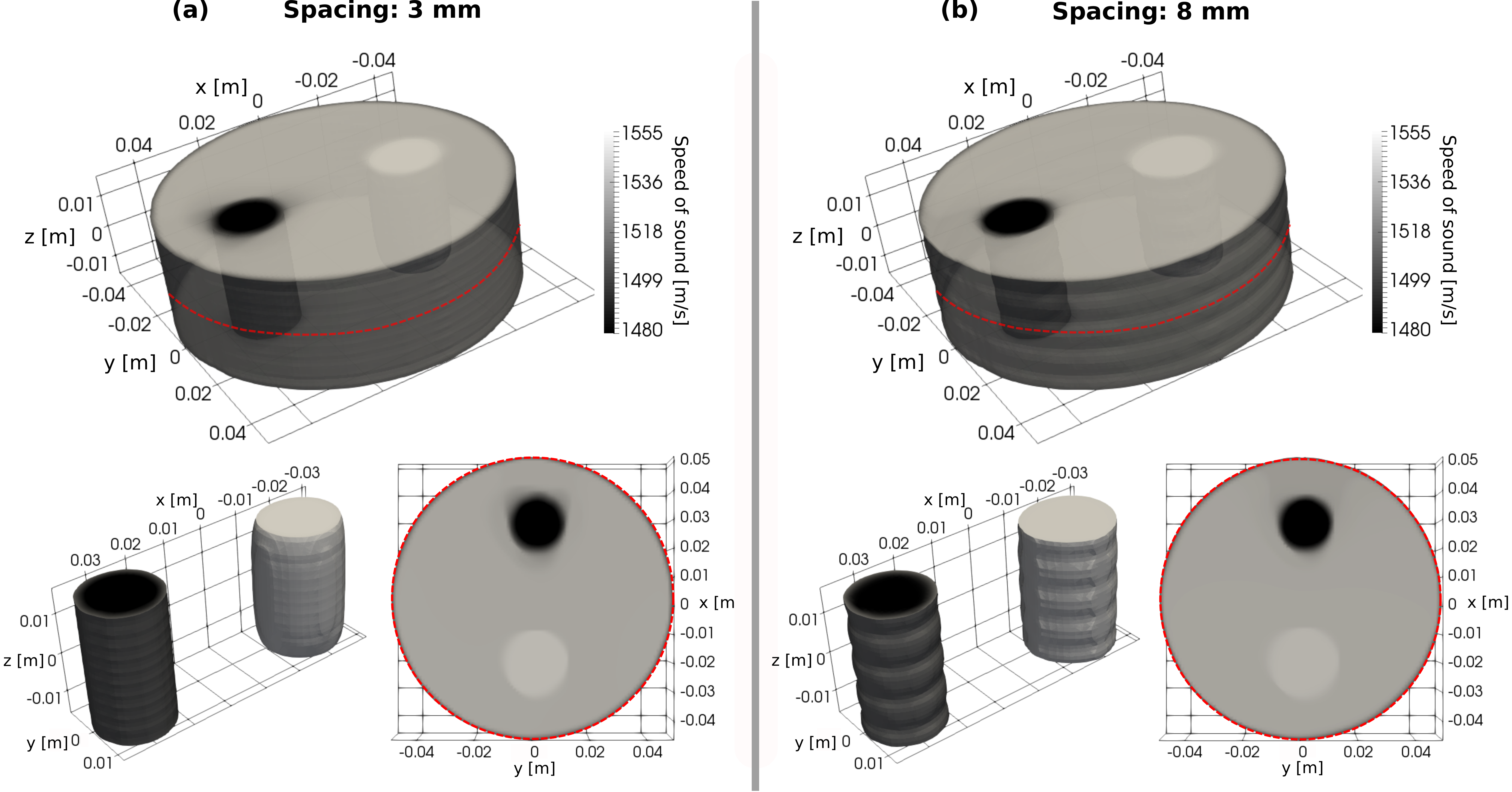}
\caption{\label{Full3D_rec} 3D velocity reconstruction for (a) 3~mm and (b) 8~mm
spacing between measurements at different elevations. We show in each case the
lateral view, the isolated view of the inclusions and the cross section at $z =
0\,\text{m}$. The red lines indicate the positions of the cross sections.
}
\end{figure*}

In our previous result, we observed a vertical resolution of approximately 8~mm.
Following this, we compare experiments using a vertical spacing of 3~mm and
8~mm. The first spacing ensures the overlapping of the sensitivities at
different elevations, and the second one only avoids gaps between them.
Fig.~\ref{Full3D_rec} shows the reconstructions for both cases, in which the ROI
is a cylinder with radius 7~cm and height 3.2~cm. For Fig.~\ref{Full3D_rec}(a)
we translate the scanning system to 11 positions in ${z \in [-2.3,
2.3]\,\text{cm}}$, and Fig.~\ref{Full3D_rec}(b) has 7 positions in ${z \in
[-2.4,2.4]\,\text{cm}}$.
In both cases, our method successfully recovers the cylindrical 3D phantom,
including both heterogeneities. Because current methods using these
acquisition systems obtain 3D breast images by stacking 2D
reconstructed slices~\cite{Duric05}, our results constitute a fundamental
advance in this context.

\begin{figure*}[!t]
\centering
\includegraphics[width=7in]{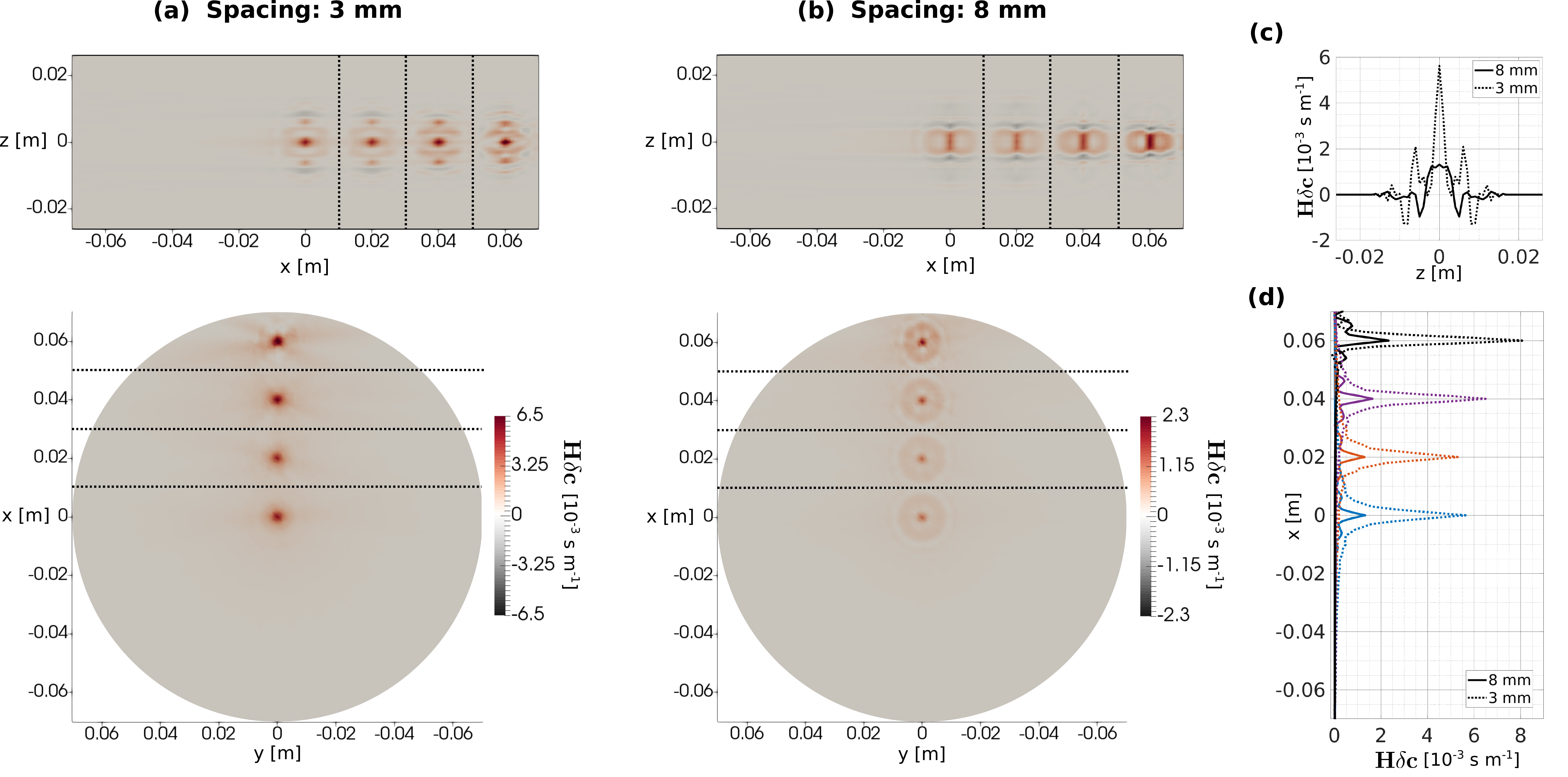}
\caption{\label{Full3D_PSF} (a), (b) Vertical and horizontal cross sections of
PSFs due to perturbations at
$(x,y,z)=(0,0,0),\,(2,0,0),\,(4,0,0),\,(6,0,0)\,\text{cm}$ for 3~mm and 8~mm
spacing, respectively.
(c) Vertical 1D cross sections of PSFs due to a
perturbation at $x = 0\,\text{m}$. (d) Horizontal 1D cross sections of PSFs
at $y = 0\,\text{m}$.}
\end{figure*}

The experiment with the largest spacing introduces oscillations in the shapes of
the recovered heterogeneities. To understand this better, we compute PSFs and
analyze differences in local resolution for both experiments, shown in
Fig.~\ref{Full3D_PSF}. As expected, the PSFs computed for 8~mm spacing are
equivalent to those already observed in Fig.~\ref{3Drec_1elev}.
However, the vertical cross sections demonstrate that the resolution length can
significantly be reduced when the sensitivities overlap. We illustrate this in
Fig.~\ref{Full3D_PSF}(c), where we plot 1D cross sections of PSFs due to
a perturbation at $x = 0\,\text{m}$. The response is remarkably sharper for 3~mm
spacing meaning that the parameters in vertical direction are better constraint.
Horizontal cross-sections also show an interesting effect, see
Fig.~\ref{Full3D_PSF}(d). PSFs for 3~mm have an increased contrast and
trade-offs that decrease in spatial extent.

\section{Discussion and conclusions}

This work presents a new transmission tomography method to estimate velocity
variations in breast tissue using ultrasound data. It minimizes
cross-correlation traveltime shifts between the observations and calibration
data in water, being internally consistent with the standard procedure of
traveltime estimations. The actual measurements are band-limited and include
finite-frequency effects of the cross-correlated pulses. Contrary to what ray
theory predicts, finite-frequency traveltimes are affected by scattering and
diffraction effects off the ray path, and their sensitivity to velocity
variations therefore extends to ellipsoidal volumes. In the context of
transmission tomography, this work represents a natural extension of ray theory
to finite-frequency waves.

Finite-frequency tomography has been extensively studied in seismology.
It is best suited for experiments with (1) velocity variations that are under
10\%~\cite{Mercerat2012} and therefore quasi-linearly related to traveltime
differences, (2) sufficient waveform similarity between cross-correlated waves,
and (3) dense data coverage. In breast imaging with USCT, all these conditions
are satisfied, therefore suggesting an ideal field of application for this
method.

Though being methodologically more involved than ray-based tomography,
finite-frequency tomography has two major advantages. First and foremost, it
correctly accounts for the frequency dependence of traveltime measurements. From
a data perspective, this means that traveltime measurements may in fact be made
in multiple frequency bands in order to capture the dispersive nature of waves
travelling through heterogeneous and dissipative media. While being beyond the
scope of this study, such multiple-frequency measurements may greatly enlarge
the dataset, thereby improving tomographic resolution \cite{Sigloch_2008}. From
an inversion perspective, the computation of correct finite-frequency
sensitivities helps to avoid inversion artifacts related to the approximation of
sensitivities. For instance, the approximation by infinitely thin rays may in
fact lead to unrealistically good resolution by virtue of the central slice
theorem \cite{Iyer_1993}. In this context, the non-zero width of
finite-frequency kernels naturally limits resolution to what is physically
possible by using waves with a certain frequency content.

The second main advantage, specifically in medical ultrasound, is the ability to
produce 3D images based on 2D acquisition systems. Slice-by-slice acquisition
devices have gained popularity due to their fast reconstructions. However, they
often suffer from artifacts caused by the 2D approximations of inherently 3D
wave phenomena~\cite{Wiskin2017,Sandhu2015,Sandhu2017}. Our work demonstrates
that finite-frequency tomography does not only offer the possibility to image
out of plane, but also to better constrain acoustic parameters by coupling data
from different slices. This is a fundamental improvement compared to the current
approaches, in which pseudo-volumetric images are built by stacking 2D coronal
slices~\cite{Duric05}. By using more accurate physical modelling, our approach
respects the finite-frequency nature of the data, thereby reducing artifacts
caused by 2D projections. Furthermore, it avoids subjective choices related to
the post-processing of reconstructed images. In this sense, this work makes the
critical contribution of providing truly 3D reconstructions using slice-by-slice
devices, which may be important to accurately locate tissue anomalies inside the
breast volume. 

A prerequisite of our method is access to calibration data that ensure
sufficient waveform similarity between cross-correlated waves. A direct
consequence of waveform similarity is the possibility to compute the Jacobian
operator at any frequency analytically, without suffering from time consuming
numerical wave propagation simulations. The resulting linear inverse problem is
very attractive for clinical practice, where fast and accurate solutions are
indispensable for the recurring experiments.
The possibility to compute properties of the forward operator prior to any
experiment reduces the time to solution significantly.

To further reduce computational cost, we develop a 1D parameterization that
allows us to represent any finite-frequency sensitivity kernel, independent of
the emitter-receiver configuration, with the same analytic function. It encodes
the full Jacobian operator, and we compute its elements on the fly for the
matrix-vector operations required in each iteration of the least-squares solver.
Our approach is ideal for GPU implementation, and, being matrix-free, it extends
very efficiently to large-scale 3D problems. The latter is probably the most
interesting application for USCT.

In addition to the tomographic method, we apply a resolution analysis based on
PSFs estimations. This is useful for a comparative assessment of the spatial
resolution and inter-parameter trade-offs that arise from different experimental
setups. For a quantitative interpretation of our results, we require better
descriptions of the observational and forward modelling uncertainties, which can
only be obtained with further experimental studies.

\section*{Acknowledgment}

The authors gratefully acknowledge fruitful discussions with Laura Ermert and
Korbinian Sager. The data used in this work is freely available in
\textit{USCT Data Challenge 2017} online platform.

\ifCLASSOPTIONcaptionsoff
  \newpage
\fi

\bibliographystyle{IEEEtran}

\bibliography{references.bib}

\end{document}